%% file: confnote.tex
\documentclass[a4paper,12pt]{article}
\input{definitions}
\makeatletter
\newcommand*\collaboration[1]{\gdef\@collaboration{#1}}
\newcommand\afterCollaborationSpace{\vskip3pt plus 2pt minus 1pt}
\makeatother

\collaboration{The Belle II Collaboration}
\input{authors}

\begin{document}
\makeatletter
\begin{titlepage}

\includegraphics[width=3cm]{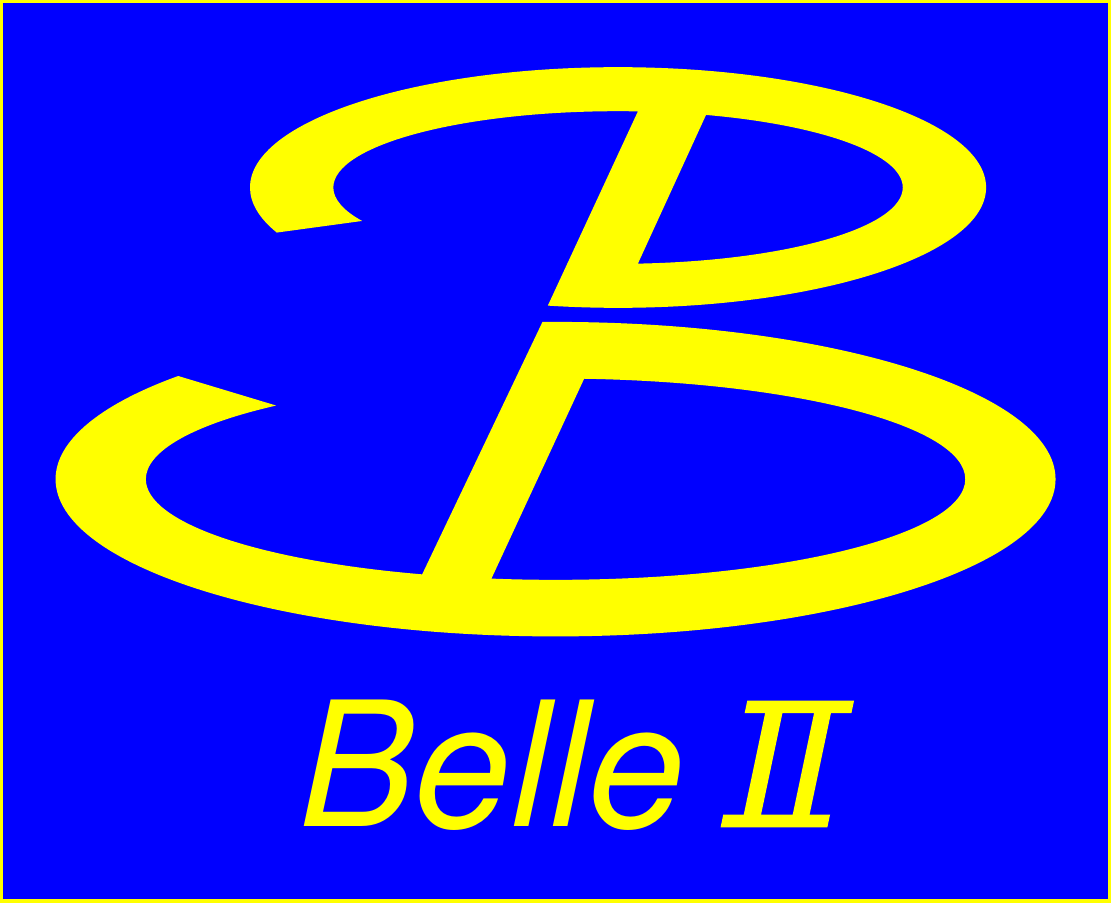}\vspace*{-3cm}

\begin{flushright}
BELLE2-CONF-PH-2022-007\\
\end{flushright}
\vspace*{3cm}

\centering
    {\LARGE \input{title}}\\[8pt]

\if!\@collaboration!\else
{\Large\@collaboration}\par
\afterCollaborationSpace
\fi
{\@author}

\begin{abstract}
    \input{abstract}
\end{abstract}

\end{titlepage}
\makeatother

\input{main}

\section*{Acknowledgements}
\input{acknowledgements}

\bibliographystyle{belle2-note}
\bibliography{references}

\end{document}

%% file: definitions.tex
\usepackage{graphicx} % Include figure files
\usepackage{epstopdf} % Allow to include eps files
\usepackage{dcolumn} % Align table columns on decimal point
\usepackage{xcolor} % Color support
\usepackage{amssymb} % Extended symbol collection
\usepackage{hyperref} % Generate pdfs with hyperlinks
\usepackage[utf8]{inputenc} % Allow UTF8 characters
\usepackage[english]{babel} % All publications are in English
\usepackage[displaymath, mathlines]{lineno} % Line numbers for drafts
\usepackage{blindtext} % For testing
\usepackage{comment}
\usepackage{cite}

\usepackage{amsmath,bm}
\usepackage{subfigure,subfigmat}
\graphicspath{{figures/}} % Use figures directory for figures

\usepackage{enumitem}  % Align table columns on decimal point
\usepackage{feynmf}    % for feynmandiagram  
\newcommand{\overbar}[1]{\mkern 1.5mu\overline{\mkern-1.5mu#1\mkern-0.1mu}\mkern 0.1mu}
\input belle2sym.tex

%% file: authors.tex
\author{F. Abudin{\'e}n, I. Adachi, K. Adamczyk, L. Aggarwal, P. Ahlburg, H. Ahmed, J. K. Ahn, H. Aihara, N. Akopov, A. Aloisio
% ...
}

%% file: title.tex
Measurement of decay-time dependent \cp violation in \ksksks using 2019--2021 Belle~II data

%% file: abstract.tex
We report a measurement of decay-time dependent \cp-violating parameters in \ksksks decays.
We use $(198.0 \pm 3.0) \times 10^6\ \BB$ pairs collected at the \FourS resonance with the Belle II detector at the SuperKEKB asymmetric-energy $e^+e^-$ collider. 
The observed mixing-induced and direct \cp violation parameters are 
$\scp = -1.86\ _{-0.46}^{+0.91}~{\rm (stat)} \pm 0.09~{\rm (syst)}$ and 
$\acp = -0.22\ _{-0.27}^{+0.30}~{\rm (stat)} \pm 0.04~{\rm (syst)}$, respectively.

%% file: main.tex
\section{Introduction}
\begin{figure}[htb]
\centering
\includegraphics[width=0.44\textwidth]{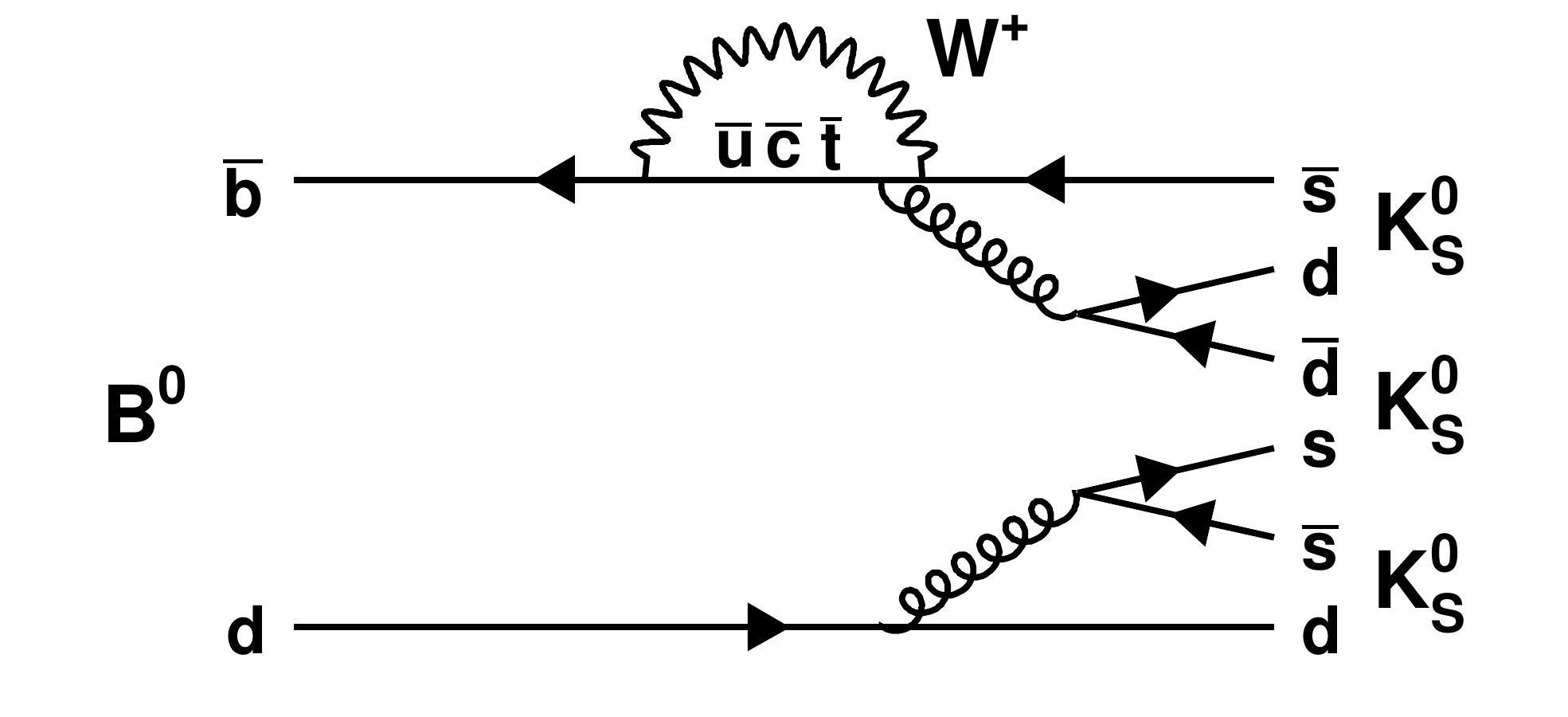}
\caption{Leading decay amplitude for the \ksksks ~decays.}
\label{fig_feynman}
\end{figure}

In the Standard Model (SM), the charmless three-body decay \ksksks is mediated by the $b \to sq\overbar{q}$ quark transition corresponding to a ``penguin''  one-loop amplitude represented by the diagram in Fig.~\ref{fig_feynman}
(charge-conjugate decays are implied hereafter unless specified otherwise).
The three-\KS final state is \cp even.
The small branching fraction $(6.0 \pm 0.5) \times 10^{-6}$\cite{Zyla:2020zbs} suppressed by the penguin loop makes this decay sensitive to a possible contribution from non-SM physics~\cite{Grossman:1996ke}.

Decay-time dependent \cp violation arises from interference between decay amplitudes with and without mixing, due to an irreducible phase in the Cabibbo-Kobayashi-Maskawa (CKM) quark-mixing matrix~\cite{Kobayashi:1973fv}.
If one of the neutral $B$ mesons produced from the $\Upsilon (4S)$ decays into a \cp eigenstate, $f_{\cp}$, at time $t_{\cp}$, and the other $B$ meson turns into a flavor-distinguishable final state,
$f_{\rm tag}$, at time $t_{\rm tag}$, the time-dependent decay rate is given by~\cite{Carter:1980hr,Carter:1980tk,Bigi:1981qs}
\begin{linenomath}
\begin{align}
\label{eqn_dt1}
\mathcal{P}(\Delta t) =   \frac{e^{-|\Delta t|/\tau_{B^0}}}{4\tau_{B^0}} (1 + q[ \mathcal{S}\sin(\Delta m^{}_d\Delta t) 
+ \mathcal{A}\cos(\Delta m^{}_d\Delta t)  ] ),
\end{align}
\end{linenomath}
where $\Delta t \equiv t_{\cp} - t_{\rm tag}$, and the \cp-violating parameters \scp ~and \acp ~are related to mixing-induced and direct \cp violation, respectively.
We refer to the $B$ meson decaying into $f_{\cp}$ as \BCP and to the other $B$ meson as \Btag.
The flavor $q$ is $+1$ ($-1$) for $B^0_{\rm tag}$ ($\overline{B}^0_{\rm tag}$), $\tau_{B^0}$ is the $B^0$ lifetime, and $\Delta m_d$ is the mass difference between the two mass eigenstates of the $B^0$-$\overbar{B}^0$ system.
The SM predicts that $\mathcal{S} \approx -\sin 2 \phi_1$ and $\mathcal{A} = 0$ in \ksksks, where $\phi_1$ is defined in terms of the CKM matrix elements as $\phi_1 \equiv \arg [-V_{cd}V_{cb}^*/V_{td}V_{tb}^*]$.
The deviation of \scp from $- \sin  2 \phi_1$ is predicted to be 0.02 with an uncertainty smaller than 0.01~\cite{Cheng:2005ug}.
The Belle\cite{Belle:2020cio} and BaBar~\cite{BaBar:2011ktx} experiments measured these asymmetries with comparable uncertainties (25\% on \scp and 20\% on \acp), where the uncertainties are dominated by the size of their \ksksks samples.
The world average values are $\scp = -0.83\pm0.17$ and $\acp = 0.15\pm0.12$~\cite{Amhis:2019ckw}.
The search for non-SM physics with a new \cp-violating phase requires additional independent measurements with improved sensitivity.

We report a measurement of \scp and \acp in \ksksks decays using a sample of $(198.0 \pm 3.0) \times 10^6$ $B\overbar{B}$ pairs collected by the Belle II experiment.
We reconstruct \ksksks decays with $\KS \ra \pipi$ as \BCP and suppress background using two boosted decision-tree (BDT) classifiers.
We then measure $q$ using the remaining charged particles in the event and \deltat from the difference between the decay positions of \BCP and \Btag.
Finally, likelihood fits are performed to determine the signal yield and \cp-violating parameters, respectively.
We use \ksksk decays as a control channel to determine fit models and to validate the fit procedure.

%==============================================
\section{The Belle II detector and data sample}
The Belle II experiment operates at the SuperKEKB asymmetric-energy $e^+e^-$ collider~\cite{Akai:2018mbz}, at KEK in Tsukuba, Japan.
The Belle II detector~\cite{Abe:2010gxa} is designed to reconstruct the final-state particles from $e^+e^-$ collisions.
Detector subsystems are arranged cylindrically around the beam interaction point.
The innermost subsystem is the vertex detector (VXD), which consists of two inner layers of silicon pixel detectors (PXD) and four outer layers of double-sided silicon strip detectors.
The second layer of the PXD is currently partially installed and covers one sixth of the design acceptance.
Compared to Belle, the larger volume of the VXD is expected to provide more acceptance for \KS vertexing. 
The main tracking device is the central drift chamber, located outside of the vertex detector.
Outside of the drift chamber, two types of Cherenkov ring-imaging detectors provide hadron identification.
The central volume is covered by a time-of-propagation detector, which uses quartz bars as Cherenkov radiator, and the forward endcap is covered by an aerogel radiator detector.
Outside of this is the electromagnetic calorimeter, which is made up of CsI(Tl) crystals.
A 1.5~T axial magnetic field is generated by a superconducting solenoid surrounding the calorimeter.
The outermost subsystem is the \KL and muon detector, which consists of iron plates interspersed with resistive plate chambers and scintillators.
Muons and \KL's are detected through their interactions with the material.
The iron plates also serve as the return yoke for the magnetic field.

We use a data set collected at the \FourS resonance in 2019--2021.
The data set corresponds to an integrated luminosity of 189.3\invfb and contains $(198.0 \pm 3.0) \times 10^6\ \BB$ pairs.
We use two types of Monte Carlo (MC) simulation samples to optimize selection criteria and to train BDTs used for event reconstruction.
In one sample $\Bz\Bzb$ pairs are generated, with one of the $B$ mesons decaying to the signal mode.
The signal MC sample is also used to determine models for the likelihood fit to the \deltat and $q$ distributions.
The other sample includes $e^+e^- \ra \qqbar \ (q=u,d,s,c), \Bz\Bzb,\ {\rm and}\ \Bu\Bub $ events that proceed with realistic hadronization and decay processes.
We use the EvtGen~\cite{Lange:2001uf} package to simulate hadron decays and KKMC~\cite{Jadach:1999vf} with Pythia~\cite{Sjostrand:2014zea} for \qqbar.
The detector response is simulated by Geant4~\cite{Agostinelli:2002hh}.
Both data and MC samples are analyzed with the Belle II analysis software framework~\cite{Kuhr:2018lps}.

%==============================================
\section{Event reconstruction}
The $\Upsilon (4S)$ is produced at the collision point with a Lorentz boost (${\beta \gamma}$) of 0.287,
and subsequently decays to $B$ and $\overbar{B}$ mesons, which are nearly at rest in the center-of-mass (CM) frame.
Therefore, the $B$ meson pairs propagate in the laboratory nearly along the boost direction with a known boost factor,
which enables us to approximate the decay-time difference between them as $\deltat = (\lboost - \tagvlboost)/\beta\gamma$.
Here, $\ell_{\cp({\rm tag})}$ is the decay position of $B_{\cp({\rm tag})}$ which is projected onto the boost axis.

%======= KS reconstruction =====================================
Pairs of oppositely charged particles with dipion mass between 457.6\mevcc and 537.6\mevcc are used to reconstruct $\KS \to \pi^+ \pi^-$ candidates.
The \KS properties are obtained from a kinematic fit of the $\pi^+$ and $\pi^-$ trajectories.
To reduce the combinatorial background from incorrectly reconstructed (``fake'') \KS candidates,
a discriminant variable is formed from a BDT classifier with 22 input variables that include kinematic quantities, particle identification information, and the number of hits in the VXD associated to the $\pi^{\pm}$ tracks, which are referred to as VXD hits.
The most discriminating variables are the angle between the directions of \KS momentum and the decay position seen from the IP in the laboratory frame, and the flight length of \KS normalized by its uncertainty.
A selection based on this BDT discriminant accepts $95\%$ of correctly reconstructed (``true'') \KS mesons and $0.59\%$ of fake \KS mesons.
Using a fit to the dipion-mass distribution in data, we confirm that the signal efficiency is consistent with MC.
Although the selection slightly biases \tauBz, it does not significantly affect \scp and \acp.
We take this effect into account as a source of systematic uncertainty.

%======= B0 reconstruction =====================================
We reconstruct $B$ candidates by combining three \KS candidates.
We select $B$ candidates using the invariant mass \mkkk and beam-energy-constrained mass $\mbc \equiv \sqrt{E_{\rm beam}^2 - |\vec{p}_{B}|^2}$, where $E_{\rm beam}$ and $\vec{p}_B$ are the beam energy and the momentum of $B$ meson in the CM frame.
The difference between the beam energy ($\sqrt{s}/2$) and the reconstructed $B$ energy in the CM frame, $\Delta E \equiv E_{\rm beam} - E_B$, is not used because of its correlation with \mbc.
We retain the $B$ candidates satisfying $5.2 < \mbc < 5.29 \gevcc$ and $5.08 < \mkkk < 5.48 \gevcc$, but exclude the candidates satisfying $5.265 < \mbc < 5.29 \gevcc$ and $5.08 < \mkkk < 5.181 \gevcc$ to avoid contamination by background due to $B^{0(+)} \ra \KS \KS K^{*0(+)}$.

The dominant source of background arises from continuum $e^+ e^- \to q\overbar{q}$ events.
We suppress the continuum background using another BDT classifier \modcs with the following input variables related to event topology:
the cosine of the angle between the thrust axes of \BCP and \Btag; the magnitude of the thrust of \Btag; the sum of the transverse momenta of the particles in the event; missing mass squared; and modified Fox-Wolfram moments~\cite{Belle:2003fgr}.
Here, the thrust axis of a $B$ meson is a unit vector $\vec{t}$ that maximizes the thrust magnitude $T \equiv (\sum_{i}\left|\vec{t}\cdot\vec{p}_i\right|)/(\sum_{i}\left|\vec{p}_i\right|)$, where $\vec{p}_i$ is the momentum of the $B$ meson $i$th daughter.
The selection on \modcs rejects 49\% of background and retains 98\% of signal.
The selection criteria for \KS candidates are determined by maximizing a figure of merit, $N_{\rm sig}/\sqrt{N_{\rm sig}+N_{\rm bkg}}$,
where $N_{\rm sig}$ and $N_{\rm bkg}$ are the yields of simulated signal and background events, respectively, satisfying $5.27 < \mbc < 5.29 \gevcc$, $5.18 < \mkkk < 5.38 \gevcc$, and $\modcs > 0.5$.

%------- chi_c0 veto ---------------------------------------------------------
In addition to the non-resonant decay amplitude, quasi-two-body decays $\Bz \ra X (\ra \KS \KS) \KS$ due to $b\ra s$ and $b \ra c$ transitions contribute to \ksksks decays.
Since we regard the decays via $b\ra s$ as signal, we veto the $b\ra c$ contribution to measure the \cp violation in a pure $b \ra s$ process.
We only expect a significant $b \ra c$ contribution from $\chi_{c0} \KS$.
The branching ratio of $\Bz \ra \chi_{c0} (\ra \KS \KS) \KS$ is around 5\% of the signal branching ratio.
We reject \Bz candidates if the invariant mass of any combination of two \KS candidates is in the range $3379 < M(\KS\KS) < 3447 \mevcc$.
This requirement rejects 90\% of the background from $\Bz \ra \chi_{c0} \KS$ and 7.5\% of signal.

\section{Measurement of $B$-meson flavor and decay-time difference}
%-------- flavor tag ---------------------------------------------------
We use a BDT-based algorithm to identify the \Btag flavor~\cite{Belle-II:2021zvj}.
It uses 13 BDTs, each of which extracts a specific signature of $b \ra c \ra s$ cascade decays from the particle identification and kinematic variables of particles not belonging to \BCP.
The outputs from the BDTs are combined by a higher-level BDT to return the value of $q$ (defined earlier) and a tagging quality variable $r$.
The variable $r$ varies from zero for no tagging information to one for unambiguous flavor assignment.
The probability density function (PDF) for signal events is represented as a modified version of Eq.~\eqref{eqn_dt1} by the probability to misidentify the flavor, $w$, and its difference between $B^0$ and $\overbar{B}^0$, $\Delta w$,
\begin{align}\label{eqn_dt3}
\mathcal{P}^{\rm TD}_{\rm sig}(\Delta t,q) & =  \frac{e^{-|\Delta t|/\tau_{B^0}}}{4\tau_{B^0}}  
 \left(\mathstrut^{\mathstrut}_{\mathstrut} 1 - q\Delta w+ (1-2w)q  
[ \mathcal{S}\sin(\Delta m^{}_d\Delta t) + \mathcal{A}\cos(\Delta m^{}_d\Delta t)  ]\right).
\end{align}
The events are classified into seven independent $r$ intervals.
For each of these intervals, $w$ and $\Delta w$ are determined using flavor-specific $B$-meson decays with large branching fractions~\cite{Belle-II:2021zvj}.
%The effective tagging efficiency considering the misidentification probabilities is $30.0\pm1.3$\%.

%------------------ vertex fit ------------------------------
To measure \deltat, we reconstruct the decay vertices of \BCP and \Btag using information about the beam interaction point (IP), which is modeled by a three-dimensional Gaussian distribution.
The \BCP vertex position is reconstructed from \KS daughter tracks and the reconstructed \BCP trajectory that originates from the IP and points toward the reconstructed \BCP momentum.
Often \KS mesons decay outside of the VXD volume resulting in less well-measured decay positions.
Hence, the \BCP vertex resolution largely depends on the number of \KS mesons that have associated VXD hits.
In the signal MC, the fractions of events where zero, one, two, and three \KS candidates have VXD hits are 0.4\%, 8.0\%, 37.7\%, and 54.0\%, respectively.
When only one \KS has VXD hits, the \BCP trajectory helps to significantly improve the \BCP vertex resolution, reducing the average vertex position uncertainty from around 270\mum to 130\mum.

We use the tracks that do not belong to \BCP to reconstruct the \Btag vertex, excluding the tracks without an associated PXD hit and those that, combined in opposite charge pairs, yield the \KS mass.
Similarly to the \BCP vertexing, we reconstruct the \Btag trajectory using the IP information and \Btag momentum, which is calculated as the difference between the beam momentum and the \BCP momentum~\cite{Dey:2020dsr}.
The \Btag trajectory is included in the vertex fit to improve the vertex resolution and reconstruction efficiency.
We use the \chisq per degree of freedom of the vertex fit $\chisq/N$ and the vertex position uncertainty $\sigma_{\ell}$ as indicators of the quality of the \deltat measurement.
The number $N_{\rm tracks}$ of tracks in the fit, typically six, determines $N$ as $N = 2N_{\rm tracks} - 1$.
We require $\chisq/N < 100$ and $\sigma_{\ell} < 500\mum$ for \Btag.

The multiplicity of \BCP candidates in a selected event is 1.06 on average.
For events with multiple candidates, we choose the one with the smallest \BCP vertex fit \chisq, which is not correlated with the true \deltat.

We apply the following selection criteria, related to the \BCP vertex and \deltat, to the remaining candidates:
one or more \KS from \BCP is associated with VXD hits; 
if a \KS daughter track from \BCP is associated with a hit in the first PXD layer (layer-1 hit), its partner from the same \KS should also have a layer-1 hit; otherwise it is likely the layer-1 hit belongs to a particle that was incorrectly associated to \BCP, degrading the \BCP vertex resolution;
the \chisq probability of the vertex fit should be larger than 0.001 for \BCP;
$\sigma_{\ell} < 500\mum$ for \BCP;
and $-30 < \deltat < 30\ps$.
We call the events passing these criteria time-differential (TD) events and the others time-integrated (TI) events.
We use the \Btag flavor information for TI events, but not the \deltat information.
Therefore, the PDF in Eq.~\eqref{eqn_dt3} is integrated over \deltat for TI events,
\begin{align}\label{eq:Psig_timeint}
\mathcal{P}_{\rm sig}^{\rm TI}(q) = \frac{1}{2} \left( 1 - q \Delta w + (1 - 2w) q\acp\frac{1}{1 + \Delta m_d^2 \tauBz^2} \right).
\end{align}

%=============================================
\section{Determination of signal yield}
We extract the signal yields for TD and TI events separately from three-dimensional likelihood fits to the unbinned distributions of \mbc, \mkkk, and \modcs.
The likelihood function accounts for two sample components, signal and background.
For the signal component, the \mbc distribution is modeled with a Gaussian function, the \mkkk distribution with the sum of two Gaussian functions, and the \modcs distribution with an asymmetric Gaussian function.
We determine the parameters for the signal shapes with fits to the signal MC sample.
We use different parameter sets for the \mkkk distribution in TD and TI events because the latter have a broader distribution due to poorly reconstructed candidates.
For the background component, the \mbc distribution is modeled with an ARGUS function~\cite{ARGUS:1990hfq}, the \mkkk distribution with a linear function, and the \modcs distribution with the sum of a Gaussian function and an asymmetric Gaussian function.

We use \ksksk decays to determine the background parameters because their kinematic properties are similar to those of the signal decay.
The background PDF shapes are confirmed to be consistent between the two decay modes using MC samples.
The fit for \ksksk gives both the corresponding yield, used as a validation of our fitting procedures, and the background PDF parameters for the \ksksks fit.

\begin{figure}
  \begin{subfigmatrix}{2}
    \includegraphics[width=0.45\columnwidth]{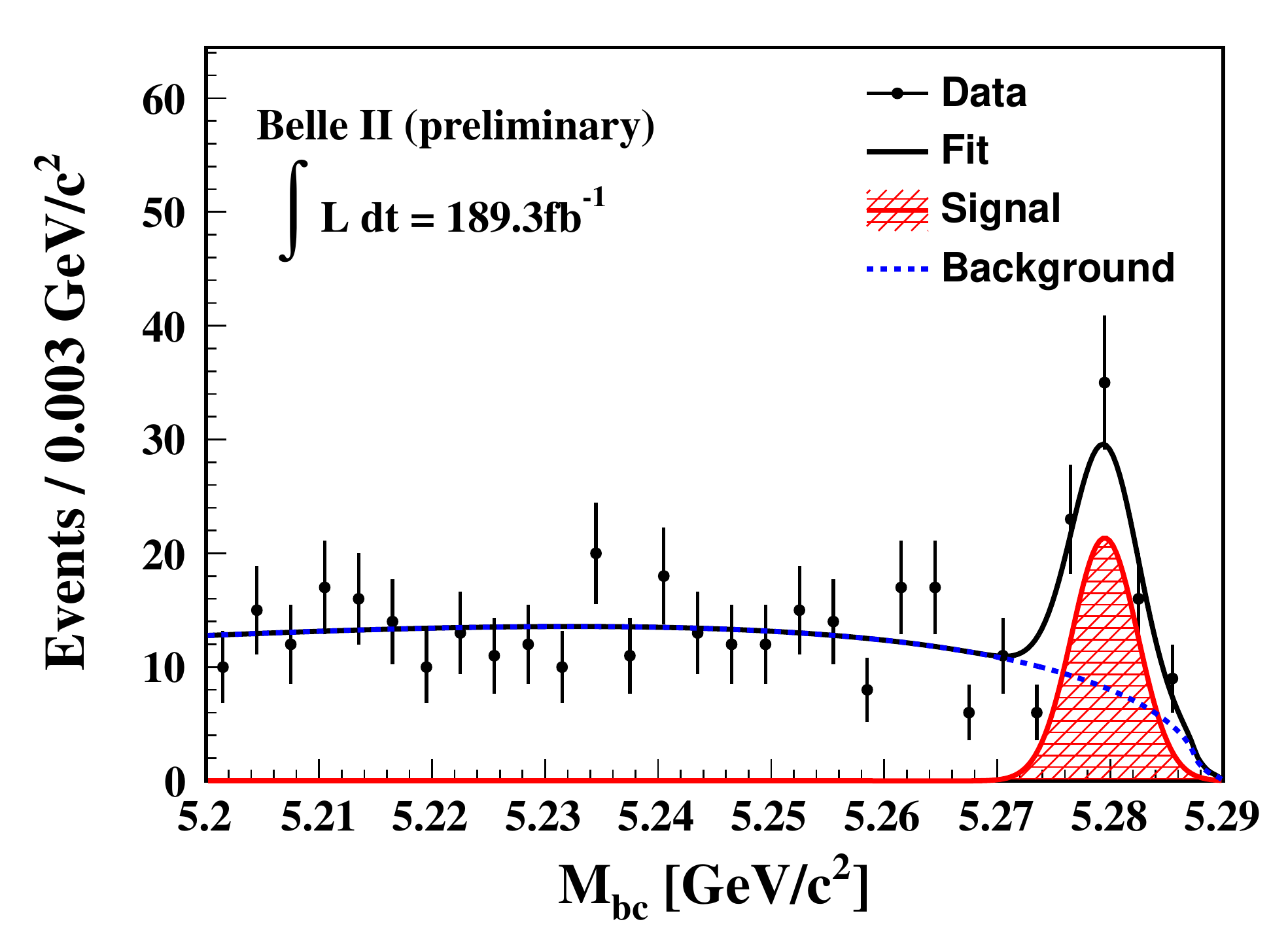}
    \includegraphics[width=0.45\columnwidth]{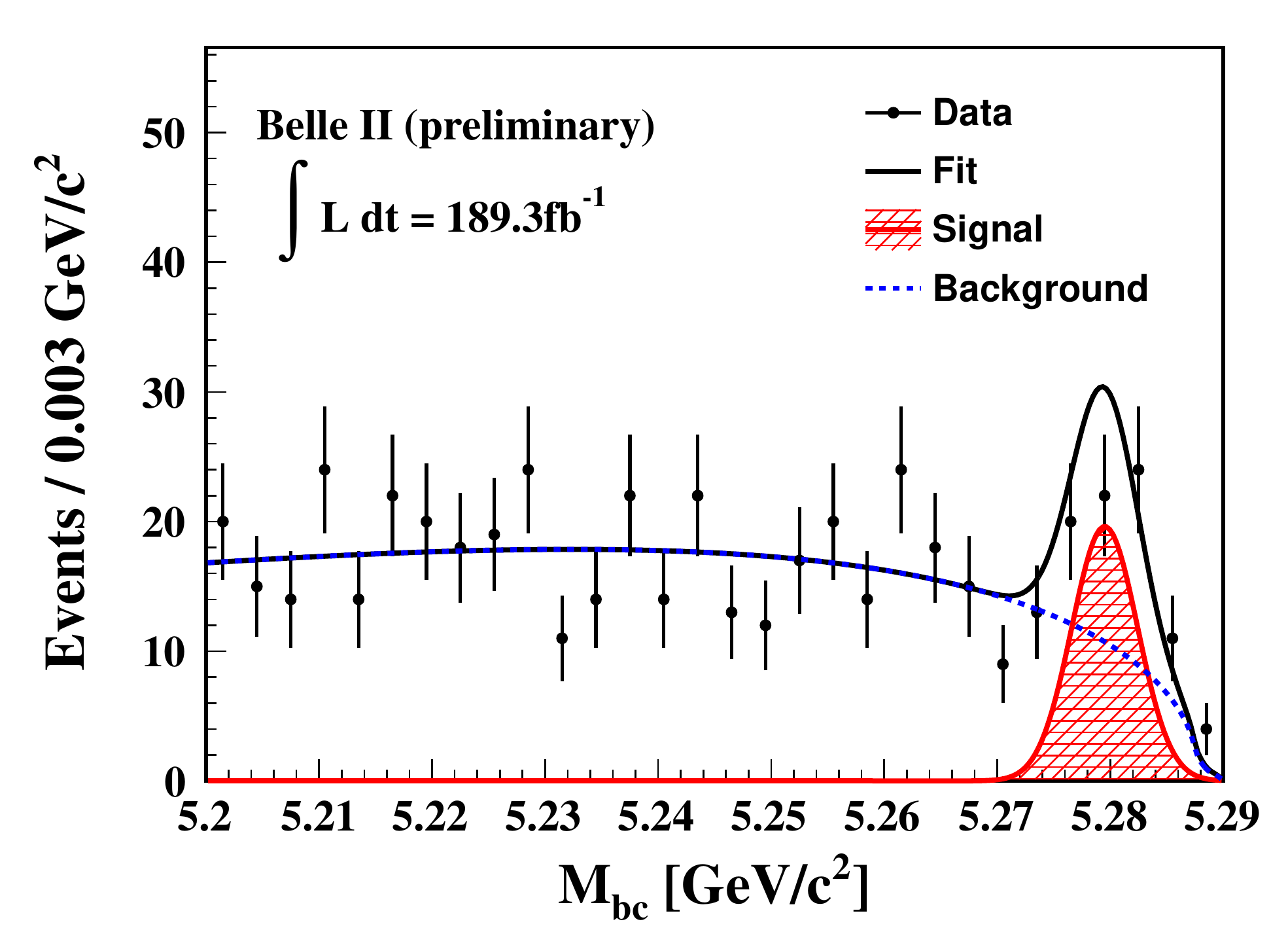}
    \includegraphics[width=0.45\columnwidth]{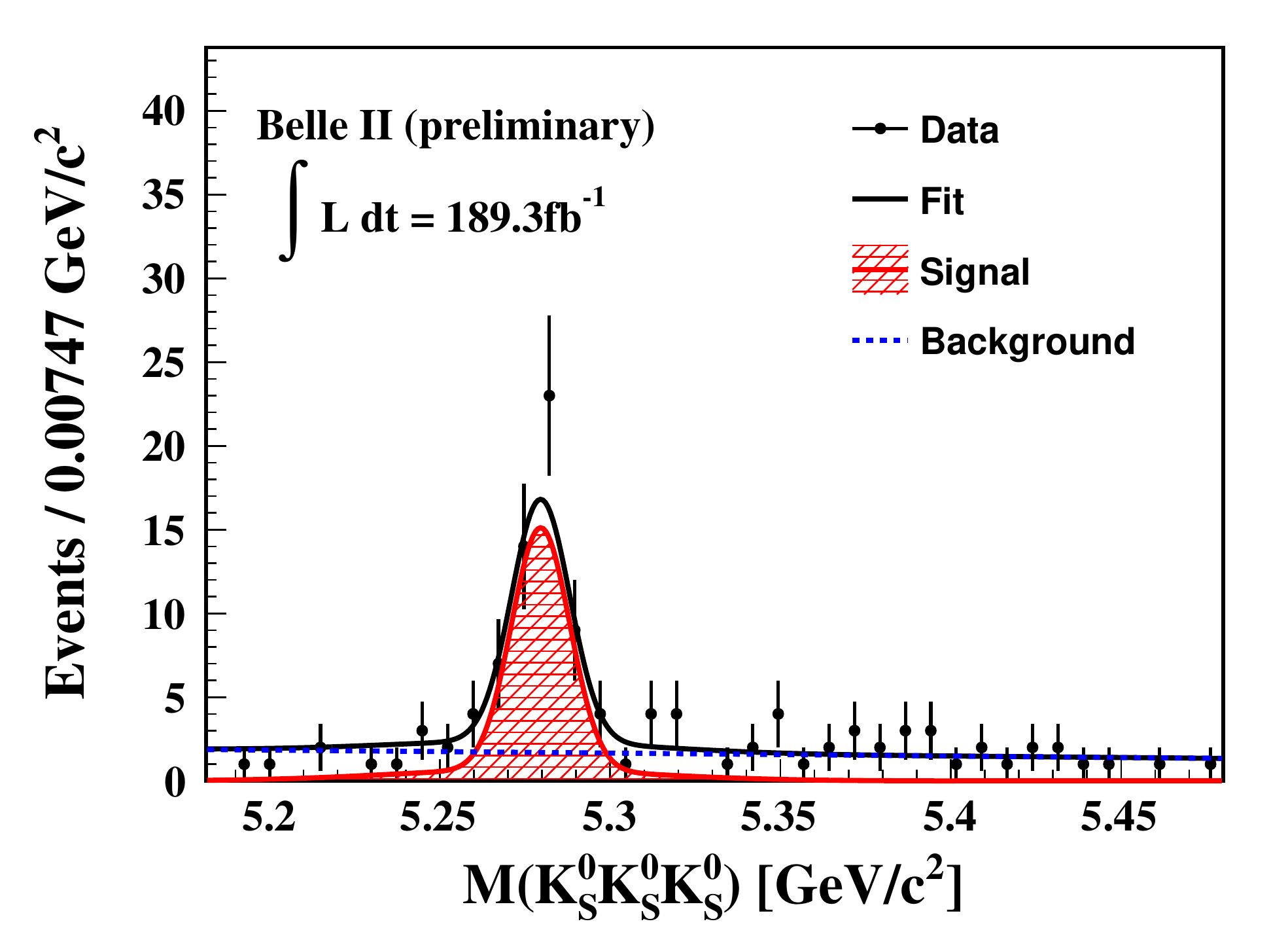}
    \includegraphics[width=0.45\columnwidth]{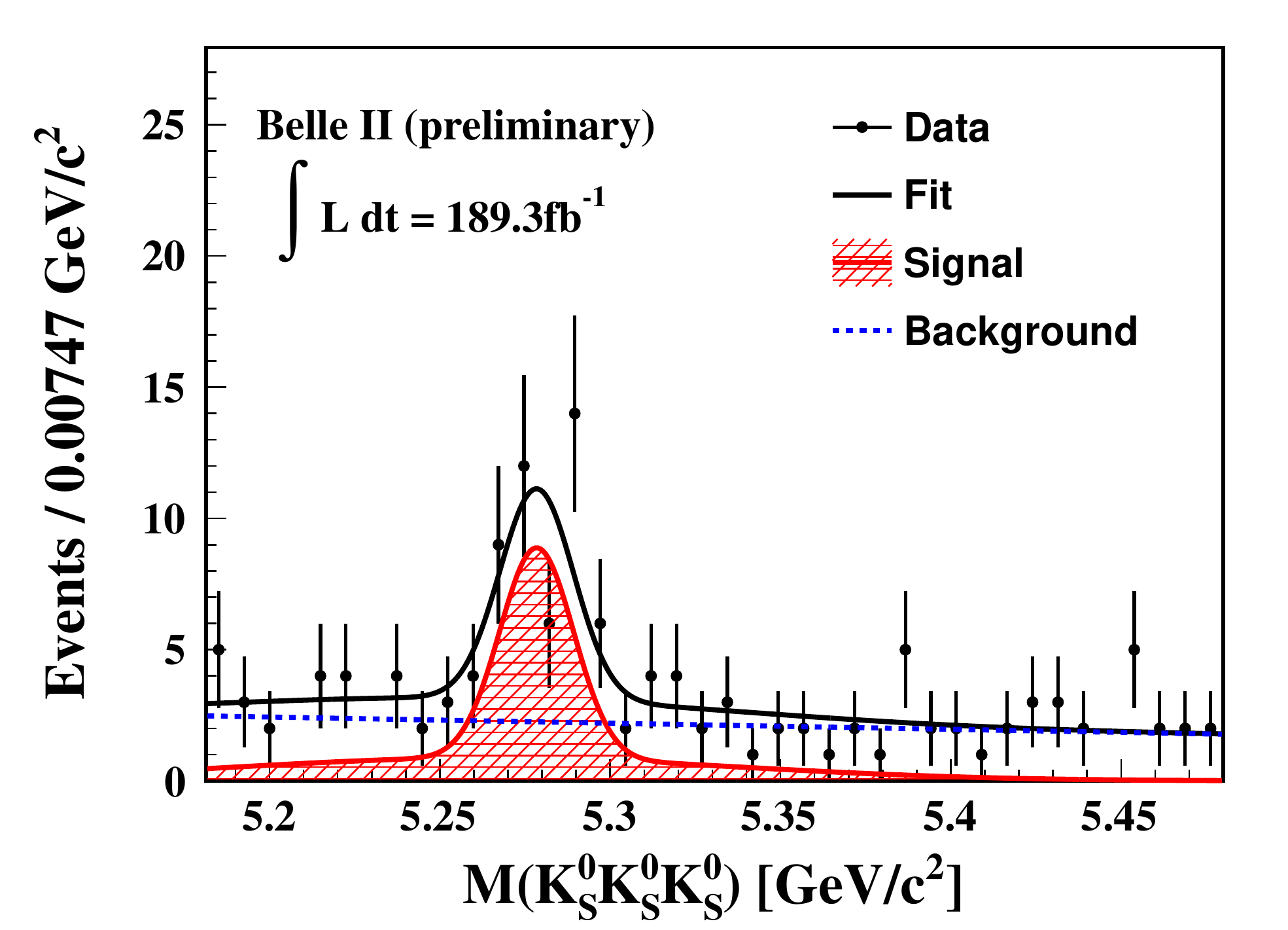}
    \includegraphics[width=0.45\columnwidth]{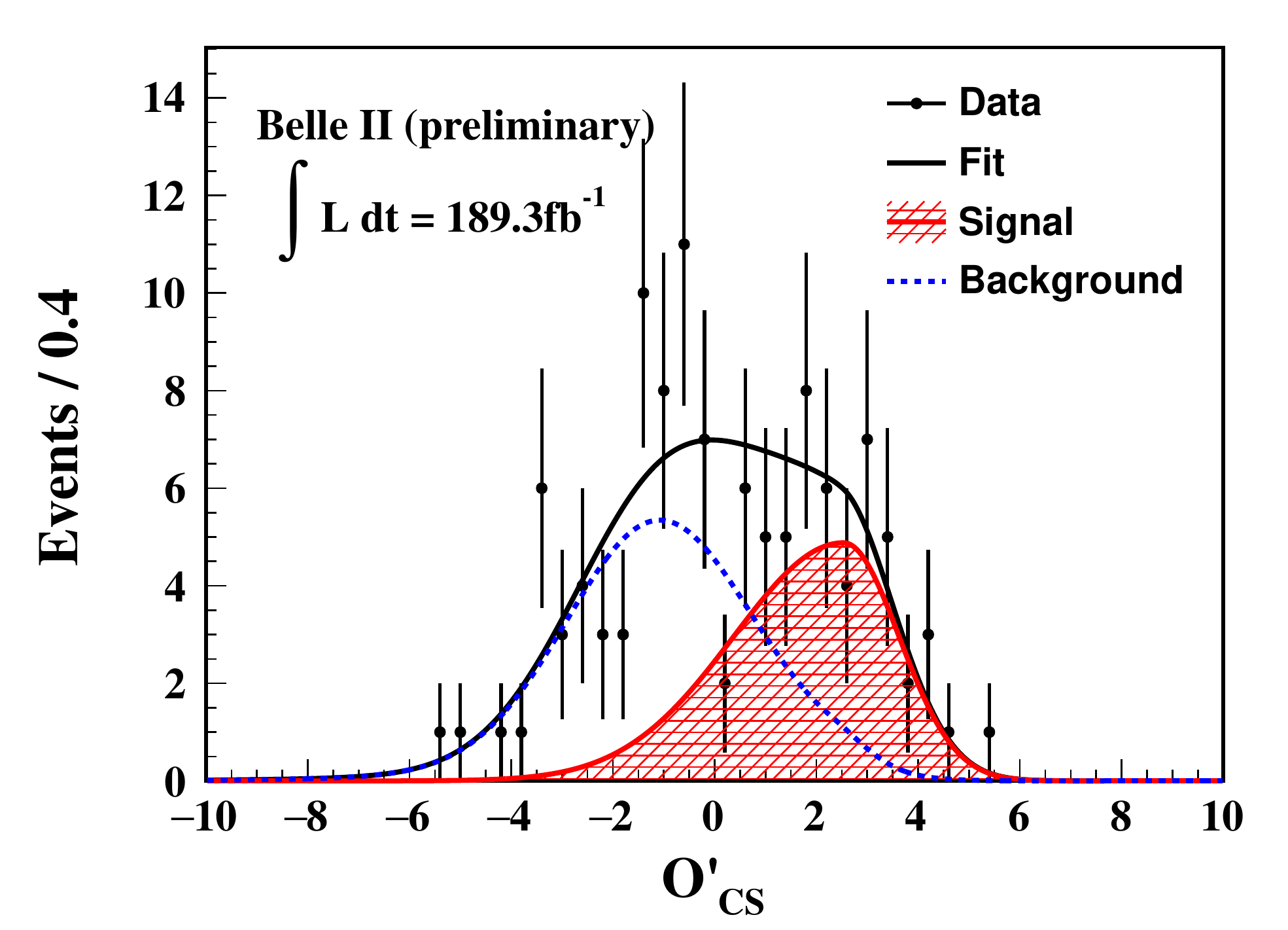}
    \includegraphics[width=0.45\columnwidth]{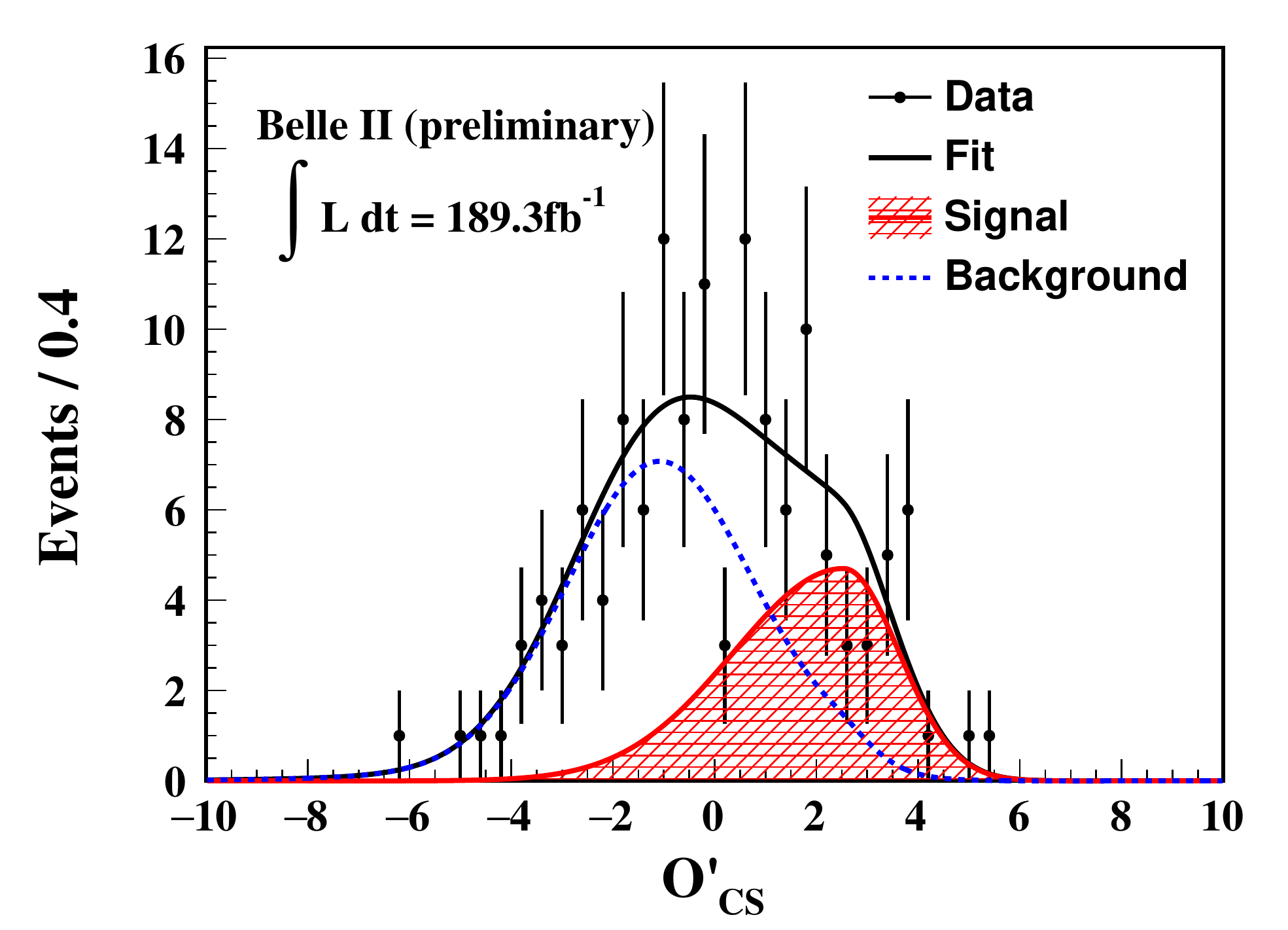}
  \end{subfigmatrix}
  \caption{Distributions of (top) \mbc, (middle) \mkkk, and (bottom) \modcs for (left) TD candidates and (right) TI candidates with fit projections overlaid.
    The \mbc distributions are restricted to events in the \mkkk signal region.
    The \mkkk and \modcs distributions are restricted to events in the \mbc signal region.}
\label{fig:cf_sigext}
\end{figure}

Figure~\ref{fig:cf_sigext} shows the results of fits to \ksksks decays, separated into TD and TI samples.
We define the signal region as $5.271 < \mbc < 5.288 \gevcc$, $5.181 < \mkkk < 5.366\gevcc$, and $-3.945 < \modcs < 5.807$ so that each range retains 99.73\% of signal TD events.
The signal yield is $53 \pm 8$ events and the purity is 54\% in the signal region for TD events, and $48_{-7}^{+8}$ events and 45\% for TI events.

%=============================================
\section{Determination of \cp-violating parameters}
We determine the \cp-violating parameters \scp~and \acp~by an unbinned maximum-likelihood fit to the \deltat and $q$ distributions combining TD and TI events restricted to the signal region.
The contribution to the likelihood function from the $j$th TD event is 
\begin{align} \label{eqn_dt4}
\mathcal{P}^{\rm TD}_j(\deltat_j, q_j) = f^{\rm sig}_j \int d(\Delta t') R(\Delta t_j - \Delta t') \mathcal{P}_{\rm sig} (\Delta t',q_j) + (1-f^{\rm sig}_{j}) \mathcal{P}_{\rm bkg} (\Delta t_j),
\end{align}
where $R(\delta \Delta t)$ is the response function of the \deltat measurement, $f_{j}^{\rm sig}$ is the signal fraction of the $j$th event, and $\mathcal{P}_{\rm bkg}$ is the $\Delta t$ distribution of background events.
The response function consists of three components: 
detector resolutions for \BCP and \Btag,
bias due to secondary tracks from charmed intermediate states for \Btag,
and a correction to the boost factors due to their small, but non-zero, CM momentum. 
The parameters for the response function are fixed to the values obtained from a fit to signal MC events.
We calculate $f_{j}^{\rm sig}$ from the signal-extraction fit including the tagging probability $r$ and the cosine of the polar angle of the \BCP momentum in the CM frame.
The additional variables are introduced to avoid fit biases due to implicitly considering equal distributions that differ across sample components~\cite{Punzi:2003wze}.
The $r$ distribution for the background component is obtained from the $\mbc < 5.265\gevcc$ sideband of the control channel.
The distribution $\mathcal{P}_{\rm bkg}$ is modeled in the same way as the response function component for the detector resolution. 
Its parameters are determined by a fit to the data sideband, $\mbc < 5.265 \gevcc$. 
For TI events, we use the likelihood function of Eq.~\eqref{eqn_dt4} integrated over \deltat,
\begin{align} \label{eqn_dt5}
\mathcal{P}^{\rm TI}_j(q_j) = f^{\rm sig}_j \mathcal{P}^{\rm TI}_{\rm sig} (q_j) + \frac{1-f^{\rm sig}_{j}}{2}.
\end{align}

To validate the analysis procedure, we reconstruct \ksksk decays without using the position information of $K^+$ in the vertex fit and extract \scp~while fixing \acp~at zero.
The result $\scp = 0.37 _{-0.33}^{+0.31}$ is consistent with no \cp violation and thus supports the robustness of our analysis procedure.
Only TD events are used in the fit for the control channel.

\begin{figure}[htb]
  \centering
  \includegraphics[width=0.45\columnwidth]{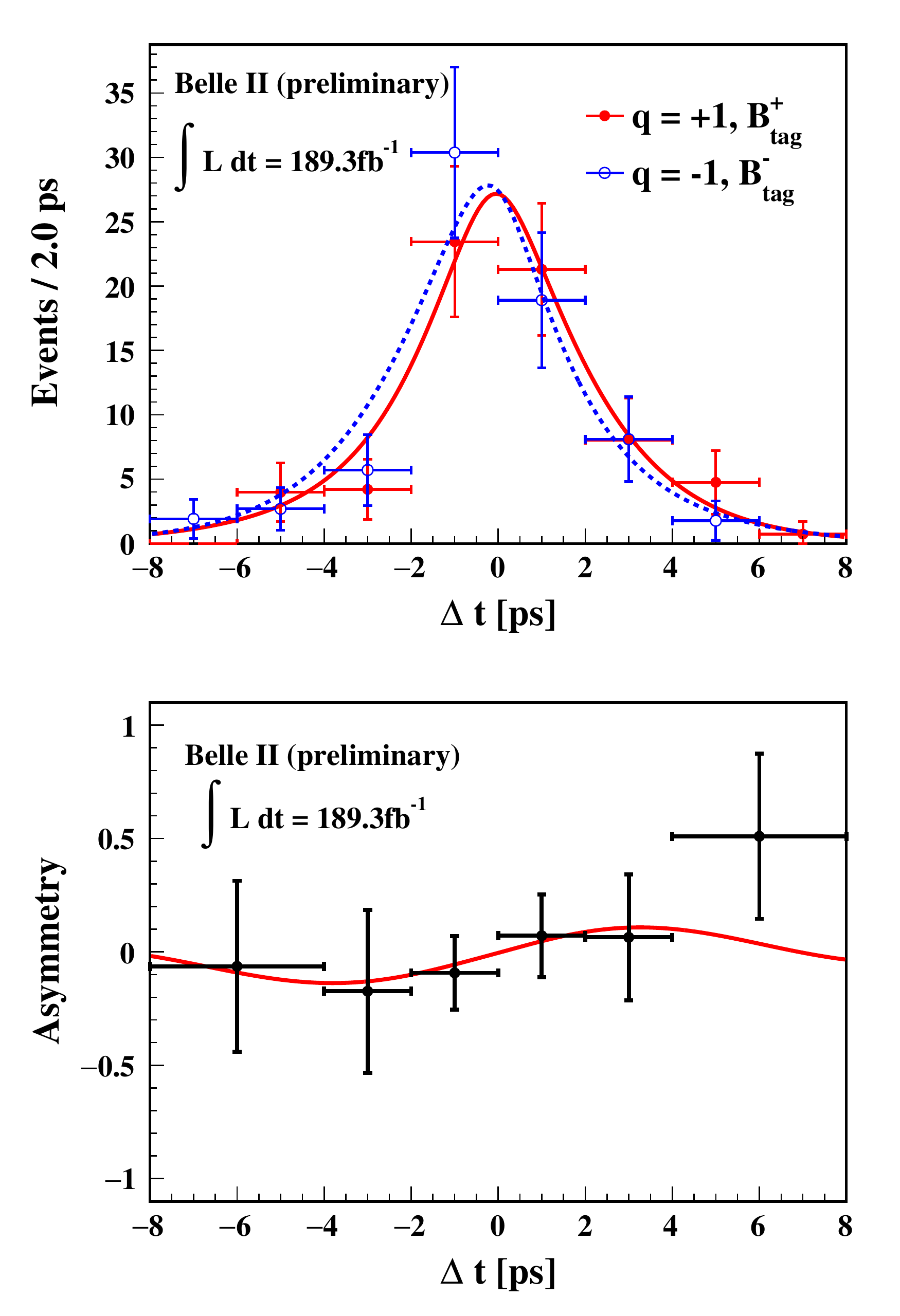}
  \includegraphics[width=0.45\columnwidth]{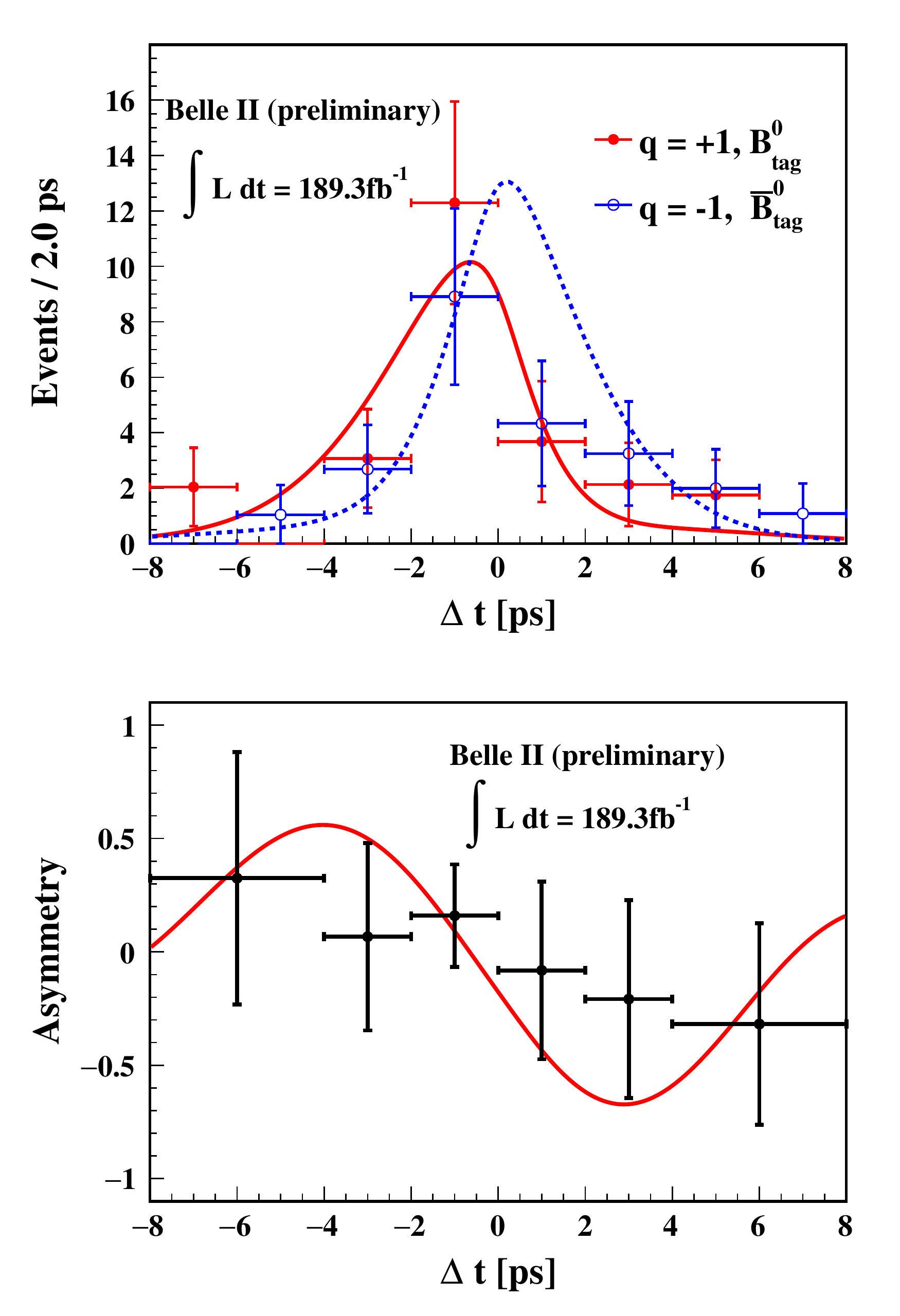}
  \caption{(top) \deltat distribution of the signal component extracted with the \sPlot technique and (bottom) its asymmetry for (left) \ksksk candidates and (right) \ksksks candidates.
    Only TD events are shown.
    In the $\Delta t$ distribution graphs, the red solid curve and filled circles represent the fit result and data for $q=+1$, 
    while the blue dashed curve and open circles represent the fit result and data for $q=-1$, respectively.
    The asymmetry is defined as $\frac{N_+(\deltat) - N_-(\deltat)}{N_+(\deltat) + N_-(\deltat)}$, where $N_{\pm}(\deltat)$ represent the number of entries with $q=\pm1$ in the corresponding \deltat bin.
    In the asymmetry graphs, the points represent data and the solid curve represents the result of the fit.
}
\label{fig_cpfit}
\end{figure}

Using \ksksks decays, we obtain $\scp = -1.86_{-0.52}^{+ 0.60}$ and $\acp = -0.22_{-0.21}^{+0.22}$, where the uncertainties are statistical.
The uncertainties are known to be underestimated by the fit due to the small sample size.
We reevaluate them using a parametric bootstrap method, in which we generate simplified simulated experiments obtained by sampling the likelihood, with the most probable \scp~and \acp~within the physical region $\mathcal{S}^2 + \mathcal{A}^2 \leq 1$ as input parameters.
We obtain the distribution of \scp and \acp from the simplified simulated experiments and define the statistical uncertainty using 16 and 84 percentiles of the distribution.
The estimated uncertainties are $_{-0.46}^{+0.91}$ for \scp and $_{-0.27}^{+0.30}$ for \acp.
Figure~\ref{fig_cpfit} shows the signal component of the \deltat distribution separated for $q=\pm1$ using an \sPlot technique~\cite{Pivk:2004ty} and the asymmetry of the distribution.
The asymmetry is defined as $\frac{N_+(\deltat) - N_-(\deltat)}{N_+(\deltat) + N_-(\deltat)}$, where $N_{\pm}(\deltat)$ represent the number of entries with $q=\pm1$ in the corresponding \deltat bin.
The plots show only TD events.

%=============================================
%          SYSTEMATIC UNCERTAINTY
\section{Systematic uncertainties}
We consider various sources of systematic uncertainties and summarize them in Table~\ref{tab_sys}.
To evaluate the systematic uncertainty on \scp and \acp related to assumptions made on parameters of the fit model,
we repeat the fit on data using alternative values of the parameters randomly sampled based on auxiliary knowledge.
This approach is used for $w$ and \dw (referred to as flavor tagging in the table), the parameters describing the resolution function, $\tau_{B^0}$ and $\Delta m_d$ (physics parameters), the parameters for the \mbc, \mkkk, and \modcs shapes (signal fraction), and the parameters for the background \deltat shape.
The widths of the resulting distributions of \scp and \acp are taken as contributions to the systematic uncertainty.
We use the world-average values and uncertainties of $\tau_{B^0}$ and $\Delta m_d$~\cite{Zyla:2020zbs} and modify the \tauBz uncertainty considering the bias due to the \KS selection.

\begin{table}[htb]
\caption{Systematic uncertainties}
\label{tab_sys}
\centering
\begin{tabular}
 {@{\hspace{0.5cm}}l@{\hspace{0.5cm}}@{\hspace{0.5cm}}c@{\hspace{0.5cm}}@{\hspace{0.5cm}}c@{\hspace{0.5cm}}}
\hline \hline
Source & $\delta \mathcal{S}$ & $\delta \mathcal{A}$ \\
\hline
Vertex reconstruction          & 0.025  &  0.022    \\         %& 0.032  &  0.040     \\
Flavor tagging                 & 0.079  &  0.030    \\         %& 0.002  &  0.004     \\
Resolution function            & 0.012  &  0.006    \\         %& 0.016  &  0.013      \\
Physics parameters             & 0.008  &  0.000    \\         %& 0.001  &  0.000    \\
Fit bias                       & 0.003  &  0.002    \\         %& 0.012  &  0.009    \\  
Signal fraction                & 0.011  &  0.007    \\         %& 0.019  &  0.008     \\
Background $\Delta t$ shape    & 0.011  &  0.001    \\         %& 0.017  &  0.002     \\
Detector misalignment          & 0.002  &  0.004    \\      %& 0.004  &  0.005     \\ 
Resolution model               & 0.001  &  0.003    \\
Tag-side interference          & 0.014  &  0.015    \\         %& 0.001  &  0.008     \\
\hline                           
Total                          & 0.087  &  0.042    \\        %& 0.046  &  0.045    \\ 
\hline \hline
\end{tabular}
\end{table}

%vertex
The systematic uncertainty due to the vertex reconstruction is determined by varying
the parameters for the IP profile and boost vector,
track requirements for the $B_{\rm tag}$ vertex reconstruction,
criteria to select TD events, and 
 correction of helix parameter uncertainties for vertexing.
We use MC samples simulated with a misaligned detector geometry to evaluate the misalignment effect.
The systematic uncertainty on the resolution model is determined by analyzing a MC sample with alternative models of \deltat response functions.
Correlations are observed between \mkkk and $\chisq/N$ for \BCP, and between \modcs and flavor tag.
For the systematic uncertainty due to the fit bias, two sets of simplified simulated experiments are generated with and without these correlations; the fits for \scp~and \acp are performed ignoring these correlations.
We take the difference between the mean value of \scp and \acp for the two sets as a systematic uncertainty.
For tag-side interference~\cite{ref:TSI}, simplified simulated experiments are generated with
and without tag-side interference and the difference is taken as a systematic uncertainty.
The systematic uncertainty is dominated by that of the flavor tagging performance owing to the limited size of the calibration sample.

%=======================================================================
\section{Conclusion}
\begin{figure}[hbt!]
  \centering
  \includegraphics[width=0.45\columnwidth]{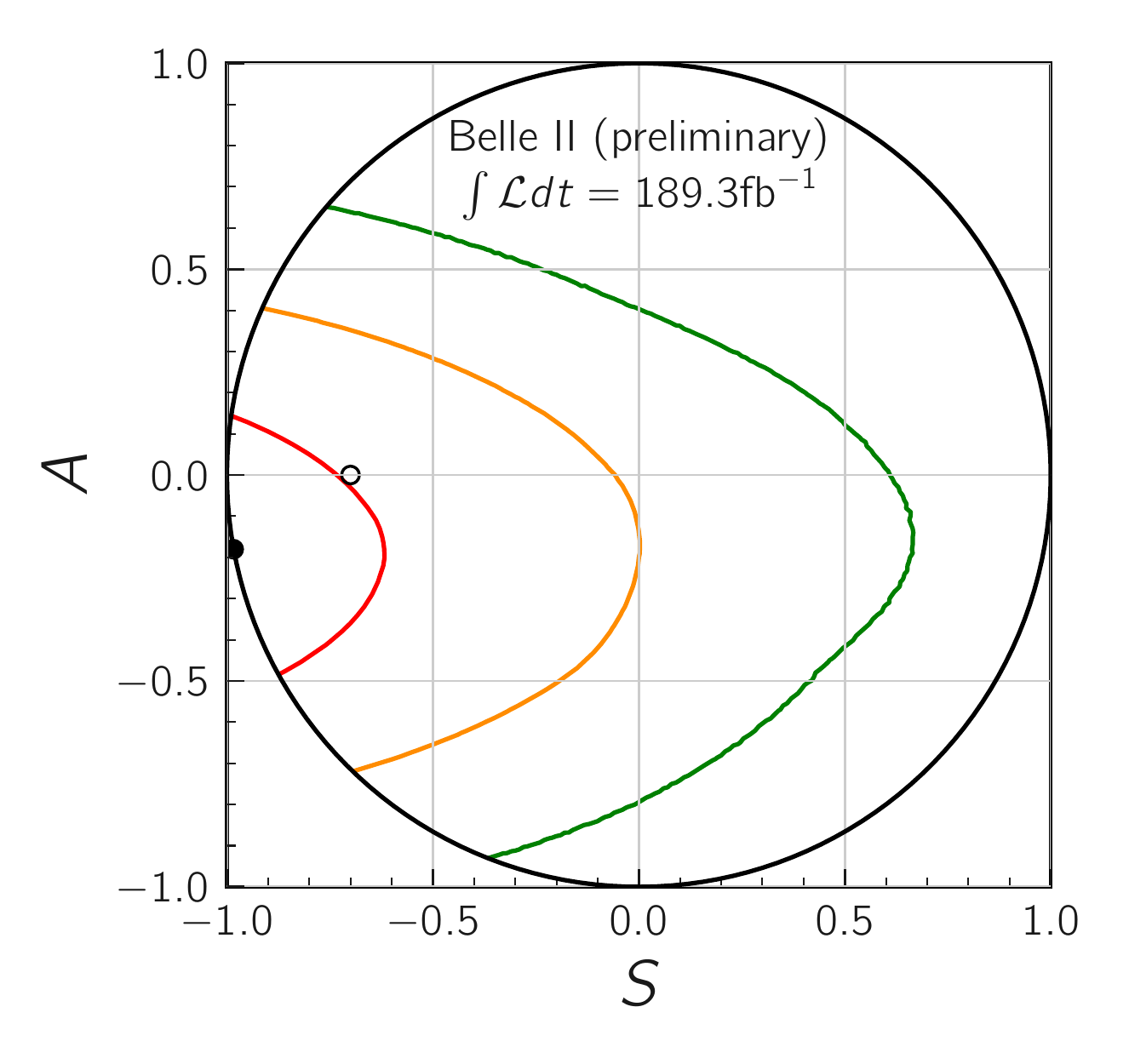}
  \caption{
    The red, orange, and green contours represent the 68.27\%, 95.45\%, and 99.73\% confidence regions for \scp and \acp given the physical constraint $\mathcal{S}^2 + \mathcal{A}^2 \leq 1$.
  The filled and open black circles indicate the most probable values for \scp and \acp in the physical region and the SM prediction $(\scp,\acp)=(-\sin2\phi_1,0)$ based on measurements in $B^0\ra (c\overline{c}) K^0$ decays, respectively~\cite{Amhis:2019ckw}.
}
\label{fig_contour}
\end{figure}

In summary, we report a measurement of decay-time dependent \cp violation in \ksksks decays using a data set corresponding to $(198.0 \pm 3.0) \times 10^6$ \BB pairs collected with the Belle II experiment.
The measured \cp-violating parameters are
\begin{align}
\scp &= -1.86\ _{-0.46}^{+0.91}~{\rm (stat)} \pm 0.09~{\rm (syst)} \ {\rm and} \\
\acp &= -0.22\ _{-0.27}^{+0.30}~{\rm (stat)} \pm 0.04~{\rm (syst)}.
\end{align}
Figure~\ref{fig_contour} shows the confidence regions based on likelihood-ratio ordering, where both the statistical and systematic uncertainties are taken into account~\cite{Feldman:1997qc}.
Here, we constrain \scp and \acp within the physical boundary, $\mathcal{S}^2 + \mathcal{A}^2 \leq 1$.
The filled and open black circles indicate the most probable values for \scp and \acp in the physical region and the SM prediction $(\scp,\acp)=(-\sin2\phi_1,0)$ based on measurements in $B^0\ra (c\overline{c}) K^0$ decays, respectively~\cite{Amhis:2019ckw}.
The results are consistent with the latest measurements at Belle and BaBar.

%% file: acknowledgements.tex
We thank the SuperKEKB group for the excellent operation of the accelerator, the KEK cryogenics group for the efficient operation of the solenoid, and the KEK computer group for on-site computing support.